\documentclass{article}

\usepackage{arxiv}

\usepackage[utf8]{inputenc} 
\usepackage[T1]{fontenc}    
\usepackage{hyperref}       
\usepackage{url}            
\usepackage{booktabs}       
\usepackage{amsfonts}       
\usepackage{nicefrac}       
\usepackage{microtype}      
\usepackage{lipsum}
\usepackage{graphicx}

\usepackage{amsthm, epsfig, amsfonts, color, caption, float}

\usepackage{tabularx, multirow, multicol}
\usepackage{appendix}
\usepackage{graphicx}
\usepackage{subcaption}
\usepackage{caption}
\usepackage{tabularx}
\usepackage{mathtools}
\usepackage{tabularx}
\usepackage{mathtools}

\usepackage{accents}

\usepackage[english]{babel}

\usepackage{listings}

\usepackage{actuarialangle}
\usepackage{actuarialsymbol}
\usepackage{actuarialangle}
\usepackage{actuarialsymbol}

\usepackage{eso-pic}
\graphicspath{ {./images/} }

\title{Dynamic Asset Pricing Theory for Life Contingent Risks}

\author{
 Patrick Ling \\
  Department of Mathematics\\
  Utah Valley University\\
  Orem, UT 84058 \\
  \texttt{Patrick.Ling@uvu.edu} \\
}

\begin{document}
\maketitle
\begin{abstract}
Although the valuation of life contingent assets has been thoroughly investigated under the framework of mathematical statistics, little financial economics research pays attention to the pricing of these assets in a non-arbitrage, complete market. In this paper, we first revisit the Fundamental Theorem of Asset Pricing (FTAP) and the short proof of it. Then we point out that discounted asset price is a martingale only when dividends are zero under all random states of the world, using a simple proof based on pricing kernel. Next, we apply Fundamental Theorem of Asset Pricing (FTAP) to find valuation formula for life contingent assets including life insurance policies and life contingent annuities. Last but not least, we state the assumption of static portfolio in a dynamic economy, and clarify the FTAP that accommodates the valuation of a portfolio of life contingent policies.
\end{abstract}


\section{Introduction}
Life contingent contracts are financial agreements where insurance benefits (indemnities) and premium payments depend on the survival of individual(s). Two financial assets are exchanged during contracting: the financial asset of insurance benefits (indemnities), and the premium payments. Both are life contingent financial assets, because cash flows of both assets are contingent upon the health (survival) \textbf{state} of individual(s). There is the problem of pricing such life contingent assets. Actuarial business is a regulated business. When actuaries compute the premiums of life contingent policies, they stick to \textbf{the Equivalence Principle}:
\begin{equation}
  E[PV(\text{benefits outgo})]=E[PV(\text{premium income})]
\end{equation}
Although this principle can be regarded as extension of the \textbf{actuarially fair pricing} concept in insurance economics, simply saying $EPV[\text{benefits outgo}]$ is the actuarially fair pricing, or the expected value of insurance indemnity $Z(X)$, or $E[Z(X)]$, is superficial. ($X$: insured loss, or financial loss due to accident; $Z$: a real function that maps loss amount $X$ to indemnity amount; $Z(X)$ is a random variable.) Such thought ignores the financial perspective of life contingent contracts. These contracts usually have very long \textbf{durations}, so it is not advisable to ignore the financial aspect of them. Without a new theoretical framework, it will also be very hard to derive certain results like recursion. Although it is possible to derive these equations thru mathematical statistics, such proof lacks any financial intuition. We deserve an alternative method that displays strong financial intuition.

We can apply the \textbf{Fundamental Theorem of Asset Pricing}  to life contingent cash flow of an asset in order to find the valuation. We use \textbf{Farkas' Lemma} to prove the FTAP in a one step model, under the assumptions of non-arbitrage, complete financial market. Then we rewrite the FTAP using the language of conditional expectation. This allows us to extend FTAP to accommodate a multiple step model. Although most financial literature points out that discounted asset prices is a martingale, we show that discounted asset price is a martingale only when no dividends are paid out under all random states. Then, with the help of ordinary differential equations, we derive valuation formulas for life insurance policies and life contingent annuities. Finally, we carefully investigate the details of information set as we price a portfolio of risky assets in a dynamic economy.

\section{Dynamic Asset Pricing Theory in a One Step Model}
\label{sec:one-step}
We first introduce the \textbf{dynamic economy}. We then define \textbf{arbitrage opportunity} in such dynamic economy. Later, we prove that lack of arbitrage opportunity implies existence of \textbf{stochastic discount factor (SDF)}. We will present you with the assumption that financial market is \textbf{complete}; thus, SDF is assumed to unique in a one step model.

\subsection{The Dynamic Economy}
Assume there are $m$ assets, whose asset prices at time $t$ is $\boldsymbol{p}_t=(p_t^1,p_t^2,\dots,p_t^m)$. Assume these $m$ assets generate dividends at time $t$ of $\boldsymbol{d}_t=(d_t^1,d_t^2,\dots,d_t^m)$. Assume an economic agent purchases a portfolio at time $t$ of $\boldsymbol{\theta}_t=(\theta_t^1,\theta_t^2,\dots,\theta_t^m)^T$. All above variables are random variables (or vectors of random variables). The cost associated with constructing the portfolio $\boldsymbol{\theta}_t=(\theta_t^1,\theta_t^2,\dots,\theta_t^m)^T$ at time $t$ is $\boldsymbol{p}_t\boldsymbol{\theta}_t=\sum_{j=1}^m p_t^j\theta_t^j$.

\subsubsection{Assumptions}

\paragraph{Asset Prices, Dividends, Portfolios are Adapted Processes}
As of time $t$, we have knowledge of $\{p_s^i\}_{s\leq t}$, $\{d_s^i\}_{s\leq t}$, $\{\theta_s^i\}_{s\leq t}$, $\{m_s\}_{s\leq t}$, and $\{a_s\}_{s\leq t}$ , because all these stochastic processes are \textit{assumed to be \textbf{adapted} in dynamic asset pricing theory}

\subsection{Asset Pricing in a Dynamic Economy}

\subsubsection{Arbitrage Portfolio}

A portfolio $\boldsymbol{\theta}_t=(\theta_t^1,\theta_t^2,\dots,\theta_t^m)^T$ is said to be an \textbf{arbitrage opportunity} if
\begin{align}
  \boldsymbol{p}_t\boldsymbol{\theta}_t&\leq 0\\
  (\boldsymbol{p}_{t+1}+\boldsymbol{d}_{t+1})\boldsymbol{\theta}_t&\geq 0
\end{align}
\dots with at least one of the inequalities to be strict with positive probability.

\subsubsection{Farkas' Lemma}

Let $\boldsymbol{A}\in \mathbb{R}^{m\times n}$ and $\boldsymbol{b}\in \mathbb{R}^{m\times 1}$ . Then exactly one of the following two statements is true:

\paragraph{Statement 1} - There exists an $\boldsymbol{x}\in \mathbb{R}^{n\times 1}$ such that $\boldsymbol{Ax}=\boldsymbol{b}$ and $\boldsymbol{x}\succeq \boldsymbol{0}$
\paragraph{Statement 2} - There exists a $\boldsymbol{y}\in\mathbb{R}^{m\times 1}$ such that $\boldsymbol{A}^T\boldsymbol{y}\succeq \boldsymbol{0}$ and $\boldsymbol{b}^T\boldsymbol{y}< 0$

\subsubsection{Payoff Matrix of a Risky Asset}

Assume there are $m$ financial assets that can be traded. Assume there are $n$ random states for the pair $(p_{t+1}^j, d_{t+1}^j)$ (for $j=1,2,\dots,m$). Define payoff matrix:
\begin{equation}
\boldsymbol{A}=\begin{pmatrix}
p_{t+1}^{1,1}+ d_{t+1}^{1,1} & p_{t+1}^{1,2}+d_{t+1}^{1,2} & \dots & p_{t+1}^{1,n}+d_{t+1}^{1,n} \\
p_{t+1}^{2,1}+ d_{t+1}^{2,2} & p_{t+1}^{2,2}+d_{t+1}^{2,2} & \dots & p_{t+1}^{2,n}+d_{t+1}^{2,n} \\
\dots & \dots & \dots & \dots \\
p_{t+1}^{m,1}+d_{t+1}^{m,1} & p_{t+1}^{m,2}+d_{t+1}^{m,2} & \dots & p_{t+1}^{m,n}+d_{t+1}^{m,n}
\end{pmatrix}
\end{equation}

Define asset cost vector
\begin{equation}
	\boldsymbol{b}=(p_t^1,p_t^2,\dots,p_t^m)^T
\end{equation}
Because asset prices $\{p_t^j\}$ is an adapted process, there is no need to model uncertainty of $\boldsymbol{b}$ as of time $t$.

\subsubsection{Characterizing Arbitrage Portfolio}

We first assume statement 2 of Farkas' Lemma to be true. Then there exists $\boldsymbol{\theta}_t=(\theta_t^1,\theta_t^2,\dots,\theta_t^m)^T$ so that $\boldsymbol{A}^T\boldsymbol{\theta}\succeq \boldsymbol{0}$, or:
	\begin{align*}
		(p_{t+1}^{1,1}+d_{t+1}^{1,1})\theta_t^1+(p_{t+1}^{2,1}+d_{t+1}^{2,1})\theta_t^2+\dots+(p_{t+1}^{m,1}+d_{t+1}^{m,1})\theta_t^m &\geq 0 \\
		(p_{t+1}^{1,2}+d_{t+1}^{1,2})\theta_t^1+(p_{t+1}^{2,2}+d_{t+1}^{2,2})\theta_t^2+\dots+(p_{t+1}^{m,2}+d_{t+1}^{m,2})\theta_t^m &\geq 0 \\
		\dots & \\
		(p_{t+1}^{1,n}+d_{t+1}^{1,n})\theta_t^1+(p_{t+1}^{2,n}+d_{t+1}^{2,n})\theta_t^2+\dots+(p_{t+1}^{m,n}+d_{t+1}^{m,n})\theta_t^m &\geq 0
	\end{align*}
	Another implication of statement 2 is $\boldsymbol{b}^T \boldsymbol{\theta}_t<0$ , or:
	\begin{equation*}
		p_t^1\theta_t^1+p_t^2\theta_t^2+\dots+p_t^m\theta_t^m <0
  \end{equation*}
	Above results can be summarized as: under all $n$ states, the portfolio $\boldsymbol{\theta}_t$ has a non-negative payoff at time $t+1$; plus, there is a negative cost associated with constructing such portfolio $\boldsymbol{\theta}_t$ at time $t$ (with probability of 1.) Such $\boldsymbol{\theta}_t$ is an arbitrage at time $t$.

\subsubsection{Assumption: no arbitrage in financial market}

We assume that there is no arbitrage opportunity in financial market. This assumption contradicts statement 2 of Farkas' lemma. Hence, statement 1 must be true. There exists $\boldsymbol{m}_{t+1}\in \mathbb{R}^{n\times 1}$ such that $\boldsymbol{A}\boldsymbol{m}_{t+1}=\boldsymbol{b}$ and $\boldsymbol{m}_{t+1}\succeq \boldsymbol{0}$ . Let $\boldsymbol{m}_{t+1}=$ $(\pi_{t+1}^1 v_{t+1}^1,\pi_{t+1}^2 v_{t+1}^2,\dots,\pi_{t+1}^n v_{t+1}^n)^T$, then:

\begin{align*}
  (p_{t+1}^{1,1}+d_{t+1}^{1,1})\pi_{t+1}^1 v_{t+1}^1+(p_{t+1}^{1,2}d_{t+1}^{1,2})\pi_{t+1}^2 v_{t+1}^2+\dots+(p_{t+1}^{1,n}+d_{t+1}^{1,n})\pi_{t+1}^n v_{t+1}^n &= p_t^1\\
  (p_{t+1}^{2,1}+d_{t+1}^{2,1})\pi_{t+1}^1 v_{t+1}^1+(p_{t+1}^{2,2}d_{t+1}^{2,2})\pi_{t+1}^2 v_{t+1}^2+\dots+(p_{t+1}^{2,n}+d_{t+1}^{2,n})\pi_{t+1}^n v_{t+1}^n &= p_t^2\\
  \dots & \\
  (p_{t+1}^{m,1}+d_{t+1}^{m,1})\pi_{t+1}^1 v_{t+1}^1+(p_{t+1}^{m,2}d_{t+1}^{m,2})\pi_{t+1}^2 v_{t+1}^2+\dots+(p_{t+1}^{m,n}+d_{t+1}^{m,n})\pi_{t+1}^n v_{t+1}^n &= p_t^m\\
\end{align*}
Or,
\[
p_t^i=E_\pi[v_{t+1}(p_{t+1}^i + d_{t+1}^i)|I_t]
\]
We call $\boldsymbol{m}_{t+1}=(\pi_{t+1}^1 v_{t+1}^1,\pi_{t+1}^2 v_{t+1}^2,\dots,\pi_{t+1}^n v_{t+1}^n)^T$ the \textbf{stochastic discount factor (SDF)} . We call $\{\pi_{t+1}\}$ a \textbf{risk neutral} or \textbf{martingale probability}; $E_\pi$ is called a \textbf{risk neutral} or \textbf{martingale expectation}.

\subsubsection{Assumption: financial market is complete}

We assume that financial market is complete; i.e., the number of linearly independent assets in the payoff matrix is equal to the number of random states of the world. Now that market is complete, the rank of the payoff matrix is $n$:
\begin{equation*}
  \begin{pmatrix}
    p_{t+1}^{1,1}+ d_{t+1}^{1,1} & p_{t+1}^{1,2}+d_{t+1}^{1,2} & \dots & p_{t+1}^{1,n}+d_{t+1}^{1,n} \\
    p_{t+1}^{2,1}+ d_{t+1}^{2,2} & p_{t+1}^{2,2}+d_{t+1}^{2,2} & \dots & p_{t+1}^{2,n}+d_{t+1}^{2,n} \\
    \dots & \dots & \dots & \dots \\
    p_{t+1}^{m,1}+d_{t+1}^{m,1} & p_{t+1}^{m,2}+d_{t+1}^{m,2} & \dots & p_{t+1}^{m,n}+d_{t+1}^{m,n}
  \end{pmatrix}
\end{equation*}
There is a unique solution for stochastic discount factor (SDF)
\begin{equation*}
  \boldsymbol{m}_{t+1}=(\pi_{t+1}^1 v_{t+1}^1,\pi_{t+1}^2 v_{t+1}^2,\dots,\pi_{t+1}^n v_{t+1}^n)^T
\end{equation*}

\subsection{Asset Pricing for Life Contingent Risks}

\subsubsection{Assumption: risk free rate and risk free asset}

For now, we assume that there is a \textbf{risk free rate} $r=e^\delta-1$ during the period between $t$ and $t+1$ . That being said, a \textbf{risk free asset} priced $p_t^0=0$ at time $t=0$ will generate a payoff of $1+r$ with probability of 1 in all states. Payoff matrix and asset cost vector are revised accordingly:
	\begin{equation}
		\boldsymbol{A}=\begin{pmatrix}
			1+r & 1+r & \dots & 1+r \\
			p_{t+1}^{1,1}+ d_{t+1}^{1,1} & p_{t+1}^{1,2}+d_{t+1}^{1,2} & \dots & p_{t+1}^{1,n}+d_{t+1}^{1,n} \\
			p_{t+1}^{2,1}+ d_{t+1}^{2,2} & p_{t+1}^{2,2}+d_{t+1}^{2,2} & \dots & p_{t+1}^{2,n}+d_{t+1}^{2,n} \\
			\dots & \dots & \dots & \dots \\
			p_{t+1}^{m,1}+d_{t+1}^{m,1} & p_{t+1}^{m,2}+d_{t+1}^{m,2} & \dots & p_{t+1}^{m,n}+d_{t+1}^{m,n}
		\end{pmatrix} \quad \quad \quad
		\boldsymbol{b}=\begin{pmatrix}
			1\\
			p_t^1\\
			p_t^2\\
			\dots\\
			p_t^m
		\end{pmatrix}
	\end{equation}

Now that there is no arbitrage in the market, there exists $\boldsymbol{m}_{t+1}\in\mathbb{R} ^{n\times 1}$ such that $\boldsymbol{A}\boldsymbol{m}_{t+1}=\boldsymbol{b}$ and $\boldsymbol{m}_{t+1}\succeq \boldsymbol{0}$ . Let $\boldsymbol{m}_{t+1}=$ $(\pi_{t+1}^1 v_{t+1}^1,\pi_{t+1}^2 v_{t+1}^2,\dots,\pi_{t+1}^n v_{t+1}^n)^T$, then:
\begin{align*}
	(1+r)\pi_{t+1}^1 v_{t+1}^1+(1+r)\pi_{t+1}^2 v_{t+1}^2 +\dots +(1+r)\pi_{t+1}^n v_{t+1}^n&=1\\
	(p_{t+1}^{1,1}+d_{t+1}^{1,1})\pi_{t+1}^1 v_{t+1}^1+(p_{t+1}^{1,2}d_{t+1}^{1,2})\pi_{t+1}^2 v_{t+1}^2+\dots+(p_{t+1}^{1,n}+d_{t+1}^{1,n})\pi_{t+1}^n v_{t+1}^n &= p_t^1\\
	(p_{t+1}^{2,1}+d_{t+1}^{2,1})\pi_{t+1}^1 v_{t+1}^1+(p_{t+1}^{2,2}d_{t+1}^{2,2})\pi_{t+1}^2 v_{t+1}^2+\dots+(p_{t+1}^{2,n}+d_{t+1}^{2,n})\pi_{t+1}^n v_{t+1}^n &= p_t^2\\
	\dots & \\
	(p_{t+1}^{m,1}+d_{t+1}^{m,1})\pi_{t+1}^1 v_{t+1}^1+(p_{t+1}^{m,2}d_{t+1}^{m,2})\pi_{t+1}^2 v_{t+1}^2+\dots+(p_{t+1}^{m,n}+d_{t+1}^{m,n})\pi_{t+1}^n v_{t+1}^n &= p_t^m\\
\end{align*}
We still assume that the market is complete; \textit{i.e.}, the rank of the payoff matrix is $n$. As a result, there is a unique solution for stochastic discount factor (SDF) . Apparently, one solution of SDF is
\begin{equation}
	\boldsymbol{m}_{t+1}=\left(\pi_{t+1}^1 \cdot \frac{1}{1+r}, \pi_{t+1}^2 \cdot \frac{1}{1+r},\dots,\pi_{t+1}^n \cdot \frac{1}{1+r}\right)^T
\end{equation}
Due to uniqueness of SDF solution, the above SDF is the only solution that solves the system of equations. This is the unique stochastic discount factor for all assets at the current time. At different times, there are different SDF's, as payoff matrix and portfolio cost vector both change over time. 

\subsubsection{{Assumption: equivalence of mortality rates and risk neutral probabilities}}

For now, we assume that the risk neutral probabilities be replaced by mortality rates.

When we analyze both mortality and investment risk, (\textit{e.g.}, pricing variable annuities), risk neutral probabilities \textit{can} appear in life insurance and pensions; and, they are not going to be mortality rates. Another example of doing risk neutral probabilities is when we study insurance-linked securities (\textit{e.g.}, catastrophe bonds).

\section{Dynamic Asset Pricing Theory in a Multiple Step Model}
\label{sec:multiple-step}

\subsection{Pricing of Risk Free Assets}

Most financial literature refer to this model as \textbf{time value of money}. Although most results are already well established, there is need to highlight some issues with differential equations used in the current literature.

\subsubsection{Continuous Discounting}

Assuming there is only 1 risk-free asset, whose contract is binding (can be enforced). Dividend payments are made continuously from time $0$ to time $t=n$, subject to payment rate function $\delta_{p,t}$. There is risk-free force of interest (FOI) that's equal to $\delta_t$ at time $t$.

Between the time $t$ and $t+\Delta t$, we can approximate the effective interest rate by
\begin{equation}
  \delta_t \times \Delta t
\end{equation}

There is only one random state of the world, so applying FTAP, we have:
\begin{equation}
  p_t=\frac{1}{1+\delta_t\times \Delta t}(\delta_{p,t}\times\Delta t+p_{t+\Delta t})
\end{equation}
\dots which can be rewritten into a differential equation
\begin{equation}
  \frac{d p_t}{dt}-p_t\delta_t=-\delta_{p,t}
\end{equation}
The solution to the ODE is the valuation formula for risk-free assets:
\begin{equation}
  p_0=\int_0^n \delta_{p,t}e^{-\int_0^t \delta_s ds}dt +p_n e^{-\int_0^n \delta_t dt}
\end{equation}

\subsubsection{Recursive Valuation Property}

We first discount dividend payments between $t_1$ and $t_2$ , and asset price on $t_2$ , back to $t_1$ :
	\begin{equation*}
		p_{t_1}=\int_{t_1}^{t_2} \delta_{p,t}e^{-\int_{t_1}^t \delta_s ds}dt+p_{t_2}e^{-\int_{t_1}^{t_2} \delta_t dt}
  \end{equation*}
Now that we know asset price at time $t_1$, we can discount this asset price, along with dividend payments between $t_0$ and $t_1$ , back to $t_0$ :
\begin{align*}
  p_{t_0}&=\int_{t_0}^{t_1} \delta_{p,t}e^{-\int_{t_0}^t \delta_s ds}dt+p_{t_1}e^{-\int_{t_0}^{t_1} \delta_t dt}\\
  &=\int_{t_0}^{t_1} \delta_{p,t}e^{-\int_{t_0}^t \delta_s ds}dt+\left(\int_{t_1}^{t_2} \delta_{p,t}e^{-\int_{t_1}^t \delta_s ds}dt+p_{t_2}e^{-\int_{t_1}^{t_2} \delta_t dt}\right)e^{-\int_{t_0}^{t_1} \delta_t dt}\\
  &=\int_{t_0}^{t_2} \delta_{p,t}e^{-\int_{t_0}^t \delta_s ds}dt+p_{t_2}e^{-\int_{t_0}^{t_2} \delta_t dt}
\end{align*}

\subsection{Pricing of Risky Assets}

\subsubsection{Assumption: adapted processes}

We first define adapted process -- a concept in probability theory and stochastic process. Then we limit the problem we solve to dynamic economies with asset-related variables that are ``adapted'' to information structure (uncertainty tree). An informal definition of \textbf{``adapted'' process} $\{X_t\}$ is that this is (1) a time-dependent sequence of random variables, where you (2) ``only know what you're supposed to know at each point time''. \textit{E.g.}, as of time $t$, you should know the value of the random variables $X_s, s\leq t$; however, you don't know what value will be realized for $X_s,s>t$. In dynamic asset pricing theory, the following processes are assumed to be ``adapted'':
	\begin{itemize}
		\item Asset prices of the $i$-th asset: $\{p_t^i\}$
		\item Stochastic discount factors (SDF): $\{m_t\}$
		\item Dividends of the $i$-th asset: $\{d_t^i\}$
	\end{itemize}
In dynamic asset pricing theory, asset prices, SDF's, and dividends at future times $s\;(s>t)$ are considered random variables as of time $t$ because their exact values are unknown as of time $t$; however, we assume that the (risk neutral) probability distribution of these random variables is known and can be modeled given the information available at time $t$ .

\subsubsection{Information Sets}

The information set $I_t$ contains all the information made available on \textit{and} before time $t$, which includes dividend amounts, asset prices, stochastic discoun factors, on and before $t$. Information set $I_t$ also includes conditional probabilities of getting into any random state of the world in future. Hence, there is information sets at different times that are subject to:
\begin{itemize}
		\item For all $0\leq s\leq t$, it must be that $I_s\subseteq I_t$
		\item For all $0\leq t_1\leq t_2\leq  \dots\leq t_n \leq \dots$, it must be that $I_{t_1}\subseteq I_{t_2}\subseteq \dots \subseteq I_{t_n}\subseteq \dots$
		\item \textbf{Tower property} -- if $I_1\subseteq I_2\subseteq \dots \subseteq I_n$ ,
		\[
			E[ E[ \dots E[X|I_n] \dots    |I_2]|I_1]=E[X|I_1]
		\]
		\item A special case is that for $t_1\leq t_2\leq \dots \leq t_n $ ,
		\[
			E[ E[ \dots E[X|I_{t_n}] \dots    |I_{t_2}]|I_{t_1}]=E[X|I_{t_1}]
		\]
		\item \textbf{``Taking out what is known''} -- if information set $I$ includes the value of $Z$,
		\[
			E[ZX|I]=ZE[X|I]
		\]
		\item If the process $\{X_t\}$ is adapted process, then
		\[
			E[X_t Y|I_t]=X_tE[Y|I_t]
		\]
\end{itemize}

Probability literature would compare information set $I_t$ to the $\sigma$-algebra or filtration made available at time $t$. I argue that $I_t$ doesn't just include the $\sigma$-algebra, as the information about past dividends specifies the exact event (within $\sigma$-algebra) that is happening at time $t$. Such event is a collection of random states of the world, and the revelation of the exact random state will happen at a future time. 

Another reason that information set is different from $\sigma$-algebra is that information set includes conditional probabilities of random states that will could be attained in future.

\subsubsection{Pricing Kernel}

We may introduce a new process -- \textbf{pricing kernel}. At time $t$ , $a_t=m_1m_2\dots m_t$. Because stochastic discount factor process $\{m_t\}$ is an adapted process, pricing kernel $\{a_t\}$ is another adapted process -- $a_t$ is known as of time $t$. For any $t$ , $a_t$ is only contingent on the state of the world -- this is because all $m_t$'s are contingent on the state. Since
\begin{equation}
		p_t=E_\pi[m_{t+1}(p_{t+1}+d_{t+1})|I_t]
\end{equation}
Multiplying $a_t$ to the two sides
\begin{equation}
		a_tp_t=a_t E_\pi[m_{t+1}(p_{t+1}+d_{t+1})|I_t]
\end{equation}
Using ``taking out what's known'' property, we have:
\begin{equation}
		a_tp_t=E_\pi[a_{t+1}(p_{t+1}+d_{t+1})|I_t]
\end{equation}

\subsubsection{``Reduced Lottery'' Method of Finding Asset Price}
By recursively applying FTAP to the multiple step model:
  \begin{align*}
    p_t&=E_\pi[m_{t+1}(p_{t+1}+d_{t+1})|I_t]\\
    &=E_\pi[m_{t+1}(E_\pi[m_{t+2}(p_{t+2}+d_{t+2})|I_{t+1}]+d_{t+1})|I_t]\\
    &=E_\pi[m_{t+1}d_{t+1}+E_\pi[m_{t+1}m_{t+2}(p_{t+2}+d_{t+2})|I_{t+1}]|I_t]\\
    &=E_\pi[m_{t+1}d_{t+1}+m_{t+1}m_{t+2}d_{t+2}+m_{t+1}m_{t+2}p_{t+2}|I_t]\\
    &\dots \\
    &=E_\pi\left[\sum_{i=t+1}^T \prod_{j=t+1}^i m_j d_i + p_T \prod_{i=t+1}^T m_i\vline I_t\right]
  \end{align*}
   Multiplying $a_t$ to the two sides, and using ``taking out what's known'' property, we have:
    \begin{equation}
      a_tp_t=E_\pi\left[\sum_{i=t+1}^T a_id_i+a_Tp_T\vline I_t\right]
    \end{equation}
  The above conditional expectations are computed using risk neutral probabilities, that are the probabilities of the reduced lottery.  As of time $t$: all random variables ($a_i, d_i, p_i;\;t<i\leq T$) are contingent on the state of the world; however, $a_i, d_i, p_i, i\leq t$ are constant (deterministic) numbers.
  
\subsubsection{``Asset Prices Are Discounted Martingales''}

Assuming there is no intermediate dividend payouts under all random states, or:
	\begin{equation}
		d_i=0\quad t<i\leq T
  \end{equation}
	The above valuation formula reduces to
	\begin{equation}
		a_tp_t=E_\pi[a_Tp_T| I_t]
  \end{equation}
	This result is often referred to as ``asset prices are discounted martingales''. In this case, the stochastic process $\{a_tp_t\}$ is a \textbf{martingale}, adapted to time $t$.

\subsection{Pricing of Life Contingent Assets}

\subsubsection{Force of Mortality}

Actuaries are interested in mortality rates. The mortality rate over one year for $(x)$ is $q_x$:
\begin{equation}
  q_x=1-p_x=1-\frac{S_0(x+1)}{S_0(x)}=-\frac{S_0 (x+1)-S_0 (x)}{S_0 (x)}	
\end{equation}

The mortality rate over an instantaneous period is called \textbf{force of mortality}. Force of mortality for $(x)$ is denoted as $\mu_x$, and equals to:
\begin{equation}
	\mu_x = \frac{-\frac{d}{dx} S_0 (x)}{S_0 (x)}	=-\frac{\lim_{\Delta x\to 0} \frac{S_0(x+\Delta x)-S_0(x)}{\Delta x}}{S_0(x)}
\end{equation}

When $\Delta x$ is small:
\begin{align*}
	\mu_x \Delta x &\approx -\frac{S_0(x+\Delta x)-S_0(x)}{S_0(x)}\\
	&=\frac{Pr[x<T_0\leq x+\Delta x]}{Pr[T_0>x]}\\
	&=Pr[T_0 \leq x+\Delta x | T_0>x]\\
	&=\actsymb[\Delta x][]{q}{x}[]
\end{align*}

\subsubsection{Pricing of Whole Life Insurance Policies}

In this section, we use such conditional probability as we apply Fundamental Theorem of Asset Pricing (FTAP) to \textbf{continuous whole life insurance policy}.
\begin{itemize}
	\item In the whole life insurance -- continuous time model,
	\item A person can either survive or pass away within a short amount of time
	\item We assume force of mortality (FOM) is subject to the function $\mu_t$
	\item We assume there is a limiting age $\omega$ (the age at which all survivors pass away)
	\item We assume that upon death, there is a one-time death benefit of amount $b_t$, paid right at the time of death
	\item $b_t$ is actual benefit amount in dollars, not payment rate
	\item We assume force of interest (FOI) is subject to the function $\delta_t$
\end{itemize}

The conditional probability of death within the short time period of $\Delta t$ is $\mu_t\times\Delta t$, and the conditional probability of survival within the same time is $1-\mu_t\times\Delta t$. Therefore, according to Fundamental Theorem of Asset Pricing,
	\begin{equation}
		p_t^1=\mu_t\cdot \Delta_t\cdot\frac{1}{1+\delta_t\Delta_t}(0+b_{t+\Delta t})+(1-\mu_t\cdot \Delta t)\cdot\frac{1}{1+\delta_t \Delta t}(p_{t+\Delta t}^1+0)
  \end{equation}

The upper scripte 1 highlights that we are modeling the asset price under state 1 -- the state of survival at all times, until limiting age $\omega$. We do not model asset price of any death state, as price is zero in any of these random states (no dividend payout in future).

The above equation can be rewritten to reveal the average rate of change of asset price w.r.t. time under state 1:
	\begin{align*}
		\frac{p_{t+\Delta t}^1-p_t^1}{\Delta t}&=\frac{p_t^1 \delta_t\Delta t-\mu_t\Delta t b_{t+\Delta t}+\mu_t\Delta t p_{t+\Delta t}^1}{\Delta t}\\
		&=p_t^1 \delta t - \mu_t b_{t+\Delta t}+\mu_t p_{t+\Delta t}^1 \\
		&\approx p_t^1 \delta_t - \mu_t b_t + \mu_t p_t^1\quad(\Delta t\to 0)
	\end{align*}
	We develop the ordinary differential equation (ODE)
	\begin{equation}
		\frac{dp_t^1}{dt} = p_t(\delta_t+\mu_t)-\mu_tb^t
  \end{equation}
	Because there is the assumption of limiting age $\omega$, the latest possible time for $b_t$ to be paid is $t=\omega$; \textit{in other words}, no dividends will be paid after $t=\omega$ under all states of the world. Hence, asset price at $t=\omega$ is zero for all states, and state 1 is no exception.
	Therefore, we develop the boundary condition
	\begin{equation}
		p_\omega^1 = 0
  \end{equation}

  By solving the ODE along with the boundary condition, we have:
  \begin{align*}
		p_t^1 &= \int_t^\omega \mu_u b_u e^{-\int_0^u \delta_s+\mu_s ds} e^{\int_0^t \delta_s + \mu_s ds} du\\
		&=\int_t^\omega \mu_u b_u e^{-\int_t^u \delta_s+\mu_s ds} du\\
		&=\int_t^\omega \mu_u b_u e^{-\int_t^u \delta_s ds} e^{-\int_t^u \mu_s ds} du
	\end{align*}

\subsubsection{Relating Force of Mortality (FOM) and Conditional Survival $\actsymb[t][]{p}{x}[]$}

From the definition of FOM, we do substitution:
	\begin{equation*}
		\mu_{x+t} = -\frac{\lim_{\Delta t\to 0}\frac{S_0(x+t+\Delta t)-S_0(x+t)}{\Delta t}}{S_0(x+t)}=-\frac{\lim_{\Delta t\to 0}\frac{S_x(t+\Delta t)-S_x(t)}{\Delta t}}{S_x(t)}=-\frac{\frac{d}{dt}S_x(t)}{S_x(t)}=-\frac{d}{dt}\ln S_x(t)
  \end{equation*}
The second equality holds because I divided $S_0(x)$ from both the numerator and the denominator. The function input for functions $\mu_{x+t}$ , $S_x(t)$ , and $S_0(x+t)$ is $t$ ; $x$ should be treated as a constant number.	Therefore,
	\begin{equation}
		\mu_{x+t}=-\frac{\frac{d}{dt}S_x(t)}{S_x(t)}=-\frac{\frac{d}{dt}[1-F_x(t)]}{S_x(t)}=\frac{f_x(t)}{S_x(t)}
  \end{equation}
	Or equivalently,
	\begin{equation}
		f_x(t)=\mu_{x+t}S_x(t)
  \end{equation}
	This result holds regardless of the value of $x$ -- even for $x=0$

\subsubsection{Valuation of Whole Life Insurance Policies}

Using Fundamental Theorem of Calculus (FTOC - Part 1),
\[
		\int_t^u \mu_s ds = -(\ln S_0(u) - \ln S_0(t))
	\]
	Therefore, asset price at time $t$ under state 1 is equal:
	\begin{align*}
		p_t^1&=\int_t^\omega \mu_u b_u e^{-\int_t^u \delta_s ds} e^{-\int_t^u \mu_s ds} du\\
		&=\int_t^\omega \mu_u b_u e^{-\int_t^u \delta_s ds} \left(\frac{S_0(u)}{S_0(t)}\right) du\\
		&=\int_t^\omega \mu_u b_u e^{-\int_t^u \delta_s ds} \actsymb[u-t][]{p}{x}[]du\\
		&= \int_0^{\omega-t} \mu_{t+x}b_{t+x}e^{-\int_t^{t+x}\delta_s ds} \actsymb[x][]{p}{t}[] dx\quad(u=t+x)\\
		&=\int_0^{\omega-t} b_{t+x}e^{-\int_t^{t+x}\delta_s ds} f_t(x) dx
	\end{align*}
	This last result is the \textbf{valuation formula for whole life insurance -- continuous time model}. It gives you asset price of whole life insurance policy at any time under state 1.

  Assuming force of interest (FOI) is constant over time (constant means it is taking the same value at all times; keep in mind, we assume force of interest is free from stochasticity throughout the entire class). Assuming death benefit is \$1 regardless of when death happens. Then the above valuation formula can be rewritten:
	\begin{equation}
		p_t^1=\int_0^{\omega-t} e^{-\delta x} f_t(x) dx=\int_0^{\omega-t} e^{-\delta x} \actsymb[x][]{p}{t}[]\mu_{t+x} dx
  \end{equation}
	We give such valuation at time $x$ an actuarial notation $\bar{A}_x$. It denotes the \textbf{expected present value (EPV) of whole life insurance with continuous payment of \$1}; \textit{e.g.},
	\begin{equation}
		\bar{A}_x=\int_0^{\omega-x}e^{-\delta t}\actsymb[t][]{p}{x}[]\mu_{x+t}dt=\int_0^\infty e^{-\delta t}\actsymb[t][]{p}{x}[]\mu_{x+t}dt
  \end{equation}
The last equality holds because $\actsymb[t][]{p}{x}[]$ drops to zero for $t\geq \omega-x$.

\subsubsection{Pricing of Life Contingent Annuities}

In this section, we use such conditional probability as we apply Fundamental Theorem of Asset Pricing (FTAP) to \textbf{annuity payable continuously}.
\begin{itemize}
	\item In the annuity payable continuously model,
	\item A person can either survive or pass away within a short amount of time
	\item We assume force of mortality (FOM) is subject to the function $\mu_t$
	\item We assume there is a limiting age $\omega$ (the age at which all survivors pass away)
	\item We assume that after death, no benefit will ever be paid
	\item We assume that as long as the insured is alive, continuous payments are made
	\item We assume that payment rate function is $\delta_{p,t}$
	\item $\delta_{p,t}$ is the payment rate, not the actual benefit amount in dollars
	\item We assume force of interest (FOI) is subject to the function $\delta_t$
\end{itemize}

According to Fundamental Theorem of Asset Pricing,
\begin{equation}
  p_t^1=\mu_t\cdot \Delta_t\cdot\frac{1}{1+\delta_t\Delta_t}(0+0)+(1-\mu_t\cdot \Delta t)\cdot\frac{1}{1+\delta_t \Delta t}(p_{t+\Delta t}^1+\delta_{p,t}\Delta t)
\end{equation}

The upper script 1 highlights that we are modeling the asset price under state 1 -- the state of survival at all times, until limiting age $\omega$. We do not model asset price of any death state, as price is zero in any of these random states (no dividend payout in future).

The above equation can be rewritten to reveal the average rate of change of asset price w.r.t. time under state 1:
\begin{align*}
  \frac{p_{t+\Delta t}^1-p_t^1}{\Delta t}&=\frac{p_t^1\delta_t\Delta t-\delta_{p,t}\Delta t+\mu_t\Delta t\cdot p_{t+\Delta t}^1+\mu_t\delta_{p,t}(\Delta t)^2}{\Delta t}\\
  &=p_t^1\delta_t-\delta_{p,t}+\mu_t p_{t+\Delta t}^1+\mu_t\delta_{p,t}\Delta t \\
  &\approx p_t^1\delta_t-\delta_{p,t}+\mu_t p_t^1\quad(\Delta t\to 0)
\end{align*}
We develop the ordinary differential equation (ODE)
\begin{equation}
  \frac{dp_t^1}{dt} = (\delta_t+\mu_t)p_t^1-\delta_{p,t}
\end{equation}
Because there is the assumption of limiting age $\omega$, the latest possible time for $b_t$ to be paid is $t=\omega$; \textit{in other words}, no dividends will be paid after $t=\omega$ under all states of the world. Hence, asset price at $t=\omega$ is zero for all states, and state 1 is no exception.
Therefore, we develop the boundary condition
\begin{equation}
  p_\omega^1 = 0
\end{equation}

By solving the ODE along with the boundary condition, we have:
\begin{align*}
  p_t^1 &= \int_t^\omega \delta_{p,u} e^{-\int_t^u \delta_s+\mu_s ds} du\\
			&=\int_t^\omega \delta_{p,u}e^{-\int_t^u \delta_s ds} e^{-\int_t^u \mu_s ds} du
\end{align*}

\subsubsection{Valuation of Continuous Life Contingent Annuities}

Using Fundamental Theorem of Calculus (FTOC - Part 1),
\[
  \int_t^u \mu_s ds = -(\ln S_0(u) - \ln S_0(t))
\]
Therefore, asset price at time $t$ under state 1 is equal:
\begin{align*}
  p_t^1&=\int_t^\omega \delta_{p,u}e^{-\int_t^u \delta_s ds} e^{-\int_t^u \mu_s ds} du\\
  &=\int_t^\omega \delta_{p,u}e^{-\int_t^u \delta_s ds} \left(\frac{S_0(u)}{S_0(t)}\right)  du \\
  &=\int_t^\omega \delta_{p,u} e^{-\int_t^u \delta_s ds} \actsymb[u-t][]{p}{t}[]du\\
  &= \int_0^{\omega-t} \delta_{p,t+x}e^{-\int_t^{t+x}\delta_s ds} \actsymb[x][]{p}{t}[] dx\quad(u=t+x)
\end{align*}
This last result is the \textbf{valuation formula for annuity payable continuously}. It gives you asset price of annuity payable continuously at any time under state 1.

Assuming force of interest (FOI) is constant over time (constant means it is taking the same value at all times; keep in mind, we assume force of interest is free from stochasticity throughout the entire class). Assuming payment rate is \$1 per year during all times (of course, when the insured is still alive). Then the above valuation formula can be rewritten:
\begin{equation}
  p_t^1=\int_0^{\omega-t} e^{-\delta x} \actsymb[x][]{p}{t}[]dx
\end{equation}
We give such valuation at time $x$ an actuarial notation $\bar{a}_x$. It denotes the \textbf{expected present value (EPV) of annuity payable continuously with payment rate of \$1 per year}; \textit{e.g.},
\begin{equation}
  \bar{a}_x=\int_0^{\omega-x}e^{-\delta t}\actsymb[t][]{p}{x}[]dt=\int_0^\infty e^{-\delta t}\actsymb[t][]{p}{x}[]dt
\end{equation}
The last equality holds because $\actsymb[t][]{p}{x}[]$ drops to zero for $t\geq \omega-x$.

\section{Dynamic Asset Pricing Theory for Static Portfolio in a Multiple Step Model}

When we price a static portfolio that consists of multiple assets, there is need to redefine the information set.

\textit{This section is being developed and will be expanded in a future revision.}

\bibliographystyle{unsrt}  


\end{document}